\documentclass{aa}
\usepackage{graphicx}
\usepackage{times}
\usepackage{natbib}

\newcommand{\kms}{\hbox{km\,s$^{-1}$}}

\newcommand{\sn}{SN\,2002kg}
\newcommand{\ms}{\hbox{M$_{\sun}$}}
\newcommand{\nii}{[N\,{\sc ii}]}
\newcommand{\ha}{H$_{\alpha}$}

\begin{document}

\title{SN\,2002kg -- the brightening of LBV V37 in NGC 2403}

\author{Kerstin Weis\inst{1}\thanks{Lise-Meitner fellow}
\and Dominik J. Bomans\inst{1}\thanks{Guest investigator of the UK 
Astronomy Data Centre}}

\offprints{K.\ Weis, 
\email{kweis@astro.rub.de}}

\mail{K.\ Weis, Bochum, Germany}

\institute{Astronomisches Institut,
Ruhr-Universit\"at Bochum, Universit\"atsstr. 150, 44780 Bochum,
Germany}

\date{Received / Accepted}

\authorrunning{K.\ Weis \& D.\ J.\ Bomans}
\titlerunning{SN\,2002kg variations of V37}

\abstract{\sn\ is a type IIn supernova,
detected in October 2002 in the nearby  spiral galaxy NGC  2403.  
We show that the position of \sn\ agrees 
within the errors with the position of the  LBV V37.
Ground based and HST ACS images however show that V37 is still 
present after the \sn\ event. We compiled a 
lightcurve of  V37 which underlines the variablity of the object, and
shows that \sn\ was the brightening 
of V37  and not a supernova. The recent brightening is not 
a giant eruption,  but 
more likely part of an  S Dor phase. 
V37 shows strong \ha +[N\,{\sc ii}] 
emission in  recent images and in the \sn\ spectrum,  
which we interprete  
as the signature of the presence of an LBV nebula. 
A  historic spectrum lacks emission, which 
may hint that we are witnessing the formation of an LBV nebula.  
\keywords{Stars: evolution -- Stars: individual: V37 -- Stars:
mass-loss -- supernovae: individual: SN2002kg}}
\maketitle

\section{Introduction}

\subsection{LBVs and their variations \label{lbvs}}

The most massive stars, above roughly 50\,\ms, pass an  unstable phase
as they turn into {\it Luminous Blue Variable\/}  (LBVs).
In  the LBV phase, a transitional phase between the main-sequence and
Wolf-Rayet state, the stars lose large amount of mass 
($> 10^{-5}$\ms\,yr$^{-1}$). As a consequence
the evolution of these stars towards cooler temperature   
is stalled and reversed \citep{1994A&A...290..819L}. The most 
massive stars seem not to enter the red supergiant phase 
(see \citet{2001A&A...366..508V} for a discussion).
The coolest location of these stars in  the 
Hertzsprung-Russell Diagram (HRD) marks the empirical {\it Humphreys-Davidson
limit} \citep[e.g.][]{1994PASP..106.1025H}.   

\begin{figure}
{\resizebox{\hsize}{!}{\includegraphics{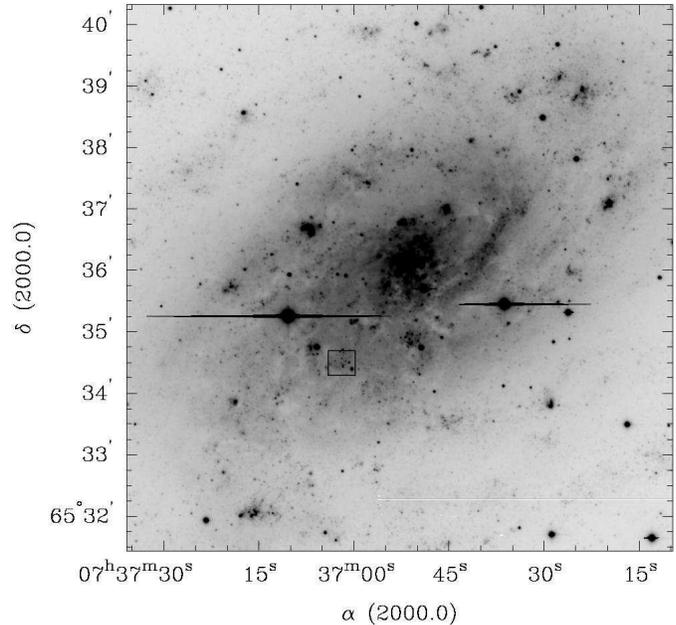}}}
\caption{A g-band image of NGC 2403 taken with the Isaac Newton telescope.
A box indicates the area in which \sn\ was found, and is shown as 
enlargement in Fig. \ref{fig:astrometrie}.}
\label{fig:ngc2403}
\end{figure}

\begin{figure*}
{\resizebox{\hsize}{!}{\includegraphics{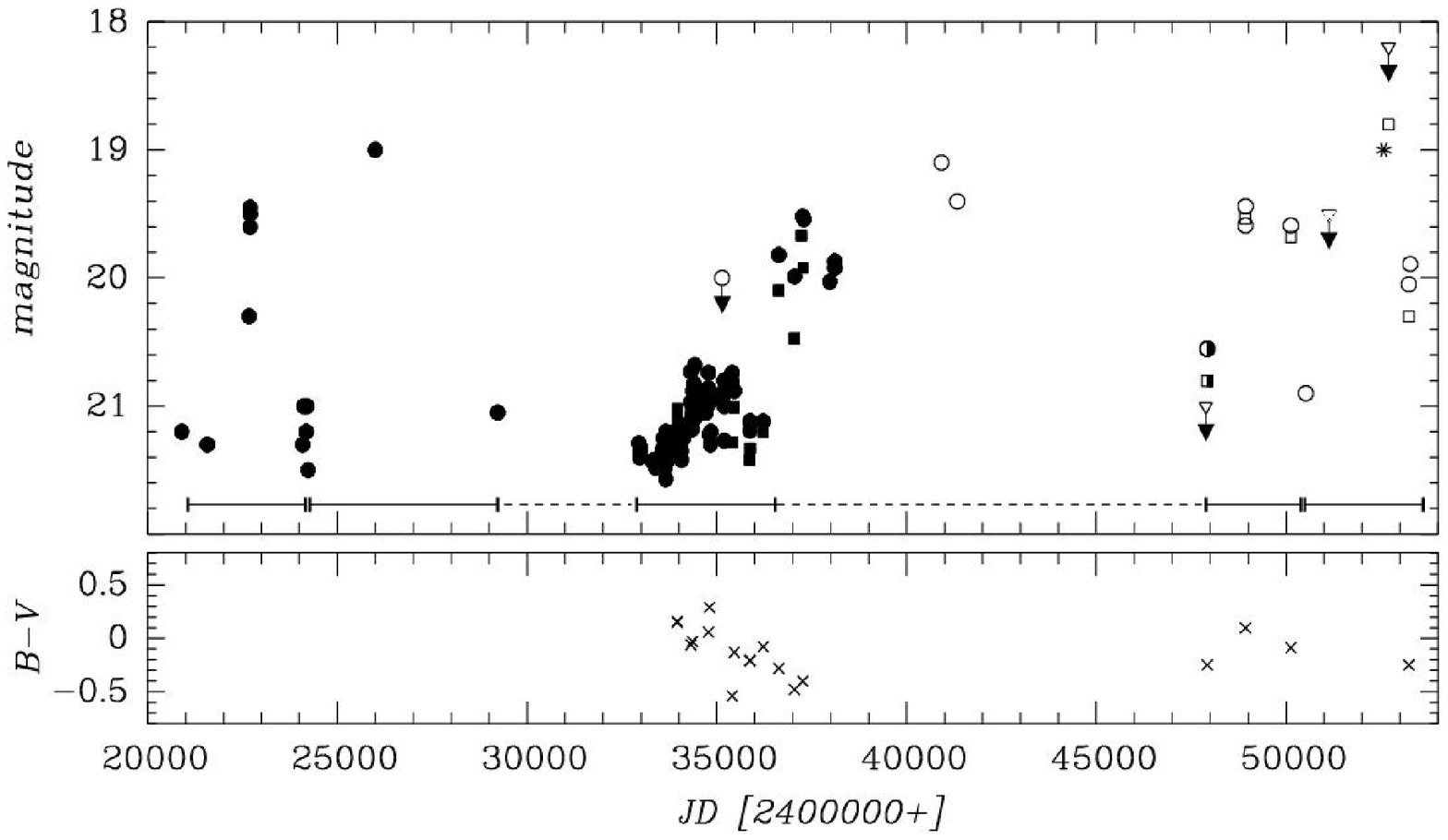}}}
\caption{The optical lightcurve of V37 (upper
panel) and  B-V color (lower panel).
Symbols are coded as: 
circles are B magnitudes,
squares are V magnitudes, triangles are R magnitudes. 
The detection of  \sn\ is indicated  with  a star, and
is roughly an R magnitude. Filled symbols  are data taken from \citet{1968ApJ...151..825T},
open symbols from this work, half-filled from
\citet{1992hdspiller}. The \sn\  detection as well as 
one R magnitude limit (hashed triangle) from \citet{2003IAUC.8051....1S}.
Possible (S)-SD periods are  indicated.
}
\label{fig:light}
\end{figure*}

LBVs show variabilities  on different timescales  and
with different strength \citep[e.g.][]{1994A&AS..106..141S},
the most classical being the S~Dor variability 
\citep[e.g.][]{1997A&AS..124..517V, 
1997A&A...318...81V, 2001A&A...366..508V}. During an S Dor
phase, which is a cyclic phenomenon of expansion and contraction at constant
luminosity, the temperature decreases and rises, respectively. 
As a consequence the spectrum changes from a hot O or B type to a cooler  
mid A or early F type \citep{1992RvMA....5....1W}, so
the colors become red and than  blue again.
It has been shown
that the  temperature variation in  such a cycle is larger for more  luminous
LBVs (amplitude-luminosity-relation, \citep{1989A&A...217...87W}).
Now, \citet{2001A&A...366..508V} subdivides the S Dor variability 
further in the  {\it short S Dor phase, (S)-SD\/} and the 
{\it long  S Dor phase, (L)-SD\/}.
The  (S)-SD is shorter than about 10 years, and 
the  (L)-SD  is  larger than 20 years.

More dramatic are the so called giant eruptions. 
These are spontaneous outbursts in  which an LBV increases  
its luminosity by several magnitudes. It stays bright for a short time
before rapidly declining back  to lower luminosities, sometimes
even lower as before the  outburst. While becoming fainter their
appearance becomes redder, presumably due to the
formation of dust \citep[e.g.][]{1999PASP..111.1124H}.
The best known example was $\eta$~Carinae's
outburst around 1843, when it was with $-1^{\rm m}$  the second
brightest star  in the southern 
hemisphere \citep{1903AnCap...9....75,1997ARA&A..35....1D}.
Other  giant eruption LBVs are P~Cygni 
($\sim$1600; \citet{1988IrAJ...18..163D}),
SN1961V in NGC 1058  (1961; \citet{1989ApJ...342..908G}) and 
SN1954J (=V12) in NGC 2403 (1954; \citet{1968ApJ...151..825T} ).
These LBVs are called 'giant eruption  LBVs' or '$\eta$ Car Variables',
or more  recently 'supernova impostors'.

The high mass loss of LBVs and the ejecta of mass during the giant
eruption leads to the formation of nebulae around LBVs 
\citep{1995ApJ...448..788N,2001RvMA...14..261W}.
These nebula are generally small ($<$ 2\,pc) but their 
expansion velocities show a  large range.
Several expand slowly ($\sim$ 30\,\kms) others can be as fast 
as  100\,\kms \citep{2003A&A...408..205W}. The fastest expansion velocities
detected are in $\eta$ Carinae with 600\,\kms (Homunculus) to
at least 2500\,\kms (outer ejecta) 
\citep[e.g.][]{2001AGM....18S0211W,2004A&A...415..595W}. 
LBV nebulae do show strong nitrogen  emission due  to
the CNO processed material \citep{1996A&A...305..229G}.

\subsection{SN2002kg and V37  in NGC\,2403 }\label{2002kg}

\sn\ was detected \citep{2003IAUC.8051....1S} in  an unfiltered 
image from October 26, 2002 with the 
{\it Katzman Automatic Imaging Telescope at 
Lick observatory (KAIT)\/}. 
\sn\ is situated in NGC\,2403 (see Fig. \ref{fig:ngc2403})
a SBc spiral in the M81 group (distance modulus = 28.14 
\citep{1988ngc..book.....T}). 
At the time of discovery \sn\ had a brightness of 19$^{\rm m}$ in the
unfiltered image.
A spectrum of the \sn\ taken  on January 6, 2003  taken with the Keck
telescope identified \sn\ as type IIn, since 
it showed for  a supernova quite narrow Balmer lines ($<$ 500\,\kms),
casting already some doubt on being a  classical supernova.
Additionally, broader components (FWHM $\sim$ 2500\,\kms) were also present.
Unusual was the detection of [N\,{\sc ii}] emission
at  6548\,\AA\  and 6584\,\AA.

V37 has first been identified as a bright blue irregular variable 
by  \citet{1968ApJ...151..825T}, 
together with V12 (alias SN1954J), V22,V35, V38, 
today confirmed LBVs \citep{1994PASP..106.1025H}. The lightcurve 
given in Fig.\ \ref{fig:light} 
shows  the  irregular behavior of V37, the original data of 
\citet{1968ApJ...151..825T} are included (filled symbols). 
From the brightness  and colors \citet{1968ApJ...151..825T} estimated 
that these stars  are F supergiants. 
A spectrum of V37 obtained in 1985 shows according 
to \citet{1987AJ.....94.1156H}
absorption-lines, a blue continuum, and no emission lines.

\begin{figure}
{\resizebox{\hsize}{!}{\includegraphics{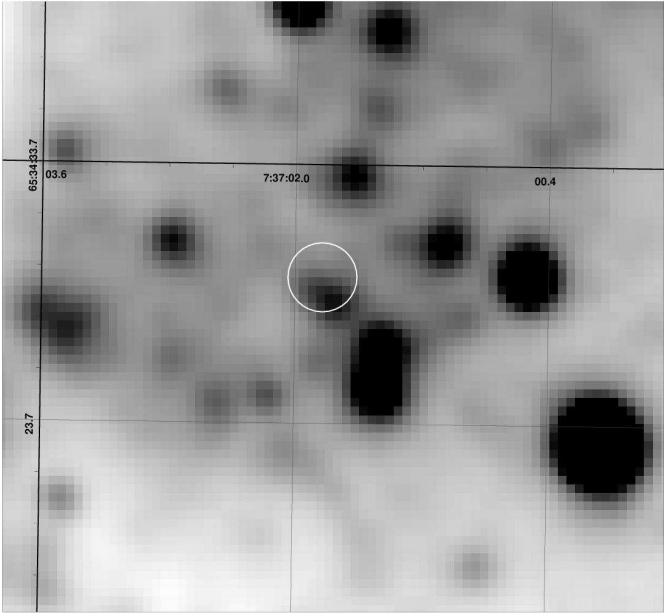}}}
\caption{This enlargement of Fig \ref{fig:ngc2403} shows the 
closer vicinity  and position (error circle) of \sn\ in NGC 2403. 
The  image was taken in with the INT in 2001, before the detection as \sn. 
}
\label{fig:astrometrie}
\end{figure}

\section{Observations and analysis}

\subsection{Astrometry}

The position of \sn\ as  given at discovery 
\citep{2003IAUC.8051....1S}  is $\alpha  =$ 7h 37m 1.83s and
$\delta =  +65\degr\,34\arcmin\,29\farcs3$ (2000.0).   
We used a deep INT archival image  in the g-band
to investigate  the surroundings of \sn. Using astrometry routines
in IRAF/STSDAS and KARMA we transferred the coordinate system of
the DSS image onto the CCD  frame.
Given an uncertainty of the process and the not corrected higher
order distortions of the CCD we reach an absolute positioning
of slightly  better than 1\arcsec.  Assuming a similar accuracy for the
position of \sn\ we generated Fig. \ref{fig:astrometrie}, 
with the g-band image in gray-scale and the error circle overlayed.
It shows clearly, that SN2002kg is coincident  with the
stellar source identified as V37 by \citet{1968ApJ...151..825T}.

\begin{figure}
{\resizebox{\hsize}{!}{\includegraphics{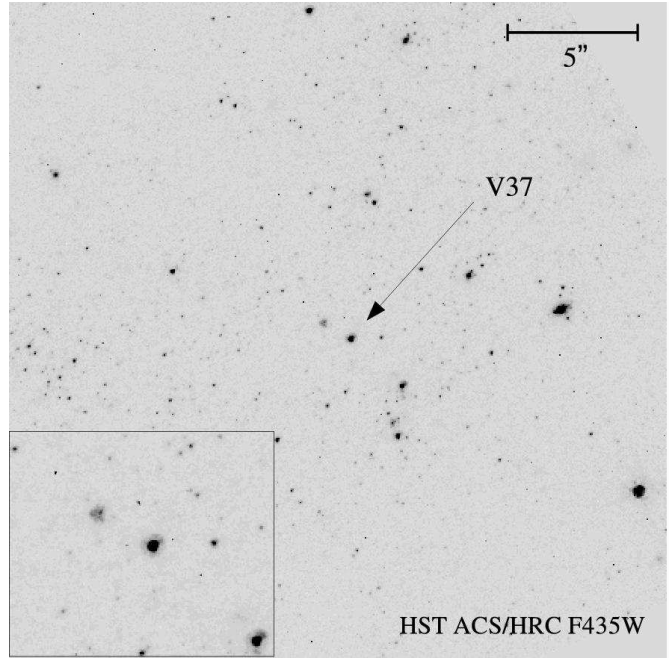}}}
\caption{HST ACS/HRC image in the F435W ($\sim$ B) image
shown the same section as in Fig. \ref{fig:astrometrie}. 
The area   of V37 is shown as enlargement in  the lower left
corner again.
}
\label{fig:acs}
\end{figure}

\subsection{Ground based photometry}

For the construction  of a lightcurve (Fig. \ref{fig:light}) we used the
measurements by \citet{1968ApJ...151..825T} as starting point.  
The B,V photometry
also provided us with a decent number of  secondary photometric
standard stars in the field of NGC 2403. We uses these stars
to tie our photometry made on a large variety of images to
Tammann-Sandage's system.  We ignore color terms between the photographic
photometry and our CCD images in  most cases. For the few very good
data we verified that the effect is generally  smaller than  the
photometric uncertainties due to crowding, seeing, and centering
of the aperture.  All measurements where done with IRAF/DAOPHOT
through Tammann-Sandage's 6\farcs7 aperture.
We measured the DSS, DSS2-blue, and DSS2-red (1955, 1997, 1989), 
as  well as two blue plates
from the Tautenburg 2\,m Schmidt (from 1970 and 1972), several
CCD images from the Isaac Newton 2.5m (1992, 1996, 2001),
Jacobus Kapteyn 1m telescopes (1996, 1998)
retrieved from the ING archive, and CCD images  
from the Tautenburg 2\,m (2003, 2004).  Whenever the object is near or below
detection limit, we also  estimated  the magnitude (or magnitude limit)
in comparison to
the stars of secondary standard sequence in the traditional way by eye.
For  comparison we also added the measurements  reported in the
discovery IAUC of \sn\ \citep{2003IAUC.8051....1S}.  
These measurements where taken
with an unfiltered CCD  mounted at the KAIT.  
We converted them into R band measurements following the recipe
described in \citet{2003PASP..115..844L}.

\subsection{HST ACS images}
Very recently HST ACS/WFC and ACS/HRC images of NGC 2403 were taken, 
which have the region V37 in the field of view.
We retrieved WFC images in the F475W, F606W, F814W, and F658N filters, 
and HRC images in the 
F435W, and F625W filters from the  HST archive.
V37 is present on the images as a  blue point source at exactly the 
position predicted based on the 2001 INT images.  This clearly shows 
that \sn\ was not a SN event, but a brightening of V37. 
We  measured  the brightness in the ACS B and V bands and added  the 
points to our lightcurve.  These  data points show  that V37 is 
fading again and  of quite blue color. 
The HRC images (Fig. \ref{fig:acs})
also reveal that next brightest (fainter by 1.6$^{\rm m}$
in F475W) object in the positional error 
circle is not a star but a blue diffuse object of uncertain nature
(see enlarged box).

\section{Discussion and conclusions}

\sn\ was from the detection  on a very strange  SN  IIn.
Being quite faint at detection (about -9$^{\rm m}$, absolute) 
and with an unusual 
spectrum it was suspected that \sn\ may not be a real
supernova. Unfortunately,  its lightcurve was not monitored. 
A comparison of ground based images (Fig. \ref{fig:astrometrie}) taken 
before the
supernova event with the position of \sn\ shows two sources, a diffuse object 
and a bright one, the latter being the LBV V37.
Astrometry on  the  ground  based  and HST ACS images showed  that 
the bright star  is indeed  coincident  with \sn.
The HST ACS images, as  well as  images  from Tautenburg show 
both sources present after 2002, which clearly  proves that \sn\  was  
rather the  brightening of a luminous star, historically known  as V37.
The diffuse source with  a size of 0\farcs18 FWHM (3.7\,pc) is too
large for being a young SNR (created 2002). The source 
is also present in our best archival images before 2002.  
It is also too faint for 
a tight cluster of massive stars, which could be progenitors of a 
type IIn supernova. The \sn\ could be a supernova in a more
distant background galaxy, but than the velocity 
of the emission lines detected \citep{2003IAUC.8051....1S}
would be different, and this is not the case.



The lightcurve of V37 in  Fig. \ref{fig:light}  shows several possible 
cycles of 2000 -- 3000 days period during the last 80 yrs as  indicated 
in Fig. \ref{fig:light}. This is consistent with (S)-SD type variations. 
However the  brightening around MJD 35000 shows a
change towards bluer colors. This is unusual as 
S Dor variabilities normally show a redward color trend.
Such an increase in brightness  with bluer colors is rarely seen 
in  LBVs, with the exceptions being presently $\eta$ Carinae 
\citep{2004MNRAS.352..447W}, 
and NGC 2366 V1, which became UV brighter during its V-band fading 
\citep{2001ApJ...546..484D}.
There is  no indication for a giant eruption  in the lightcurve of 
V37, 
but there are  still large time gaps which in theory could 
accommodate such an event in the past.

\begin{figure}
{\resizebox{\hsize}{!}{\includegraphics{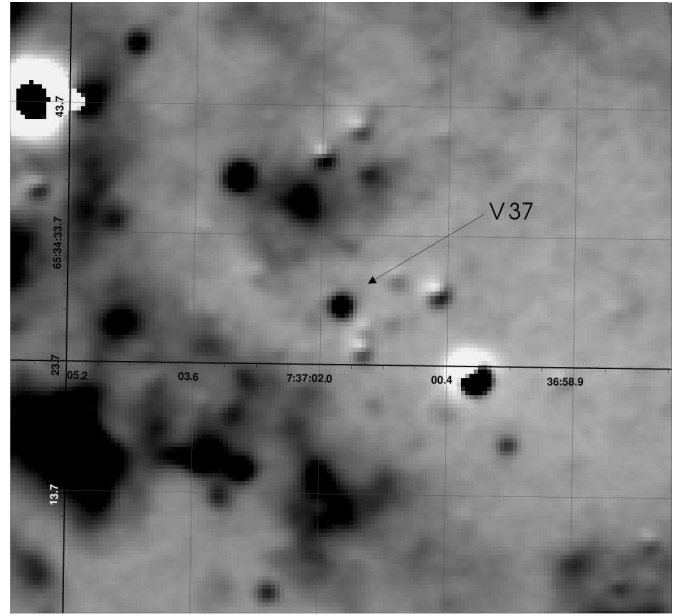}}}
\caption{This \ha\  emission line image generated from INT images 
taken in 2001, indicates that V37 is an \ha\  bright source.  
}
\label{fig:ha}
\end{figure}

With our  data  sets we  could generate two continuum corrected
\ha\ images of the region around V37. 
In  2001 V37 is a  bright 
\ha\  emitter (see Fig.\  \ref{fig:ha}). 
The HST ACS image from 2004 taken in the 
F658N filter also shows bright emission at the  position of V37 after
correction for the continuum emission.
The spectrum of \sn\ (section \ref{2002kg})  
indicates that \ha\ and  \nii\  emission 
is present at that time. One possible explanation is the creation
of an LBV  nebula  around V37  coinciding with the recent bright phase. 
Such a nebula would show strong nitrogen emission  
and could expand with velocities
as high as  detected in the \sn\  spectrum, see section \ref{lbvs}.
The H\,{\sc ii} region [H83]\,177 identified earlier
by \citet{1983AJ.....88..296H} on a  deep  \ha+\nii\ plate  
seems to coincide  within the errors
with the  position of V37, too. That could indicate that the
star has been emitting \ha\ and maybe  \nii\ as early 
as the  1980s, which would point at an earlier nebula formation.  
Still, missing emission lines in the spectrum of \citet{1987AJ.....94.1156H} 
could  mean, that the \ha\
emission would just indicate a strong stellar and  variable 
\ha\  line. The missing \nii\ lines in that spectrum may imply that the 
nebula has not  been formed in 1985.
Whether there is nebula formation during earlier times 
or not, may  only be answered with the detection (or high quality 
non-detection) of \nii\ lines
in other  historic spectra.  In any case, it seems, that 
V37 has a nitrogen enhanced nebula now.

In V37 we may witness currently the  creation of  an  LBV nebula, but it
is not yet clear whether  this nebula was created during 
a  giant  eruption, as shell ejection during several SD periods 
as in P Cyg \citep[e.g.][]{2001A&A...376..898M}, or recently 
in connection with the brightening in 2002. 
Determining the exact evolutionary  state of V37 is therefore of 
importance for 
our understanding of the LBV phenomenon and very massive stars in general.

\begin{acknowledgements}

KW is supported by the state of North Rhine-Westphalia 
(Lise-Meitner fellowship).
We thank S. Klose for kindly providing his Tautenburg 2m images, and 
H. Meusinger for scanning the historic Tautenburg plates.
We thank the referee A.M. van Genderen for his comments that helped to 
significantly improve the  paper. 
This research is partially based on data from the ING Archive.
Based partly on observations made with the NASA/ESA Hubble Space Telescope,
obtained from the data archive at the Space Telescope Institute. 
This research has made use of NASA's Astrophysics Data System.
\end{acknowledgements}

\end{document}